\documentclass[10pt,preprint2]{aastex}

\received{2010 June~7}
\revised{2010 September~20}
\accepted{2010}
\paperid{AJ/358815/ART/176005}
\cpright{}{2010}

\shorttitle{Long Wavelength Demonstrator Array Transient Sky}
\shortauthors{Lazio et al.}

\begin{document}

\title{Surveying the Dynamic Radio Sky with the Long Wavelength Demonstrator Array}

\author{T.~Joseph~W.~Lazio\altaffilmark{1,2,9}, Tracy
	E.~Clarke\altaffilmark{1}, W.~M.~Lane\altaffilmark{1},
	C.~Gross\altaffilmark{1}, N.~E.~Kassim\altaffilmark{1}, 
	P.~S.~Ray\altaffilmark{3}, D.~Wood\altaffilmark{4},
	J.~A.~York\altaffilmark{5}, A.~Kerkhoff\altaffilmark{5},
	B.~Hicks\altaffilmark{1}, E.~Polisensky\altaffilmark{1},
	K.~Stewart\altaffilmark{1}, N.~Paravastu
	Dalal\altaffilmark{6}, A.~S.~Cohen\altaffilmark{7}, \& W.~C.~Erickson\altaffilmark{8}}

\altaffiltext{1}{Remote Sensing Division, Naval Research Laboratory, 4555
	Overlook Ave.~\hbox{SW}, Washington, DC  20375  USA}

\altaffiltext{2}{NASA Lunar Science Institute, NASA Ames Research
	Center, Moffett Field, CA  94035  USA}

\altaffiltext{3}{Space Science Division, Naval Research Laboratory, 4555
  Overlook Ave.~\hbox{SW}, Washington, DC  20375-5382  USA}

\altaffiltext{4}{Praxis, Inc., 5845 Richmond Highway, Suite 700, Alexandria, VA
	22303  USA}

\altaffiltext{5}{Applied Research Laboratories, The University of Texas at
  Austin,  P.{}O.~Box~8029, Austin, TX 78713-8029  USA}

\altaffiltext{6}{American Society for Engineering Education, Washington, DC 20036  USA}

\altaffiltext{7}{The Johns Hopkins University Applied Physics Laboratory,
  11100 Johns Hopkins Road, Laurel, MD  20723  USA}

\altaffiltext{8}{School of Mathematics and Physics, University of Tasmania, Churchill Ave., Sandy Bay, Tasmania 7005 Australia}

\altaffiltext{9}{Current address: Jet Propulsion Laboratory, M/S
	138-308, 4800 Oak Grove Dr., Pasadena, CA  91109  USA; Joseph.Lazio@jpl.nasa.gov}

\begin{abstract}
This paper presents a search for radio transients at a frequency
of~73.8~MHz (4~m wavelength) using the all-sky imaging capabilities of
the Long Wavelength Demonstrator Array (LWDA).  The LWDA was a
16-dipole phased array telescope, located on the site of the Very
Large Array in New Mexico.  The field of view of the individual
dipoles was essentially the entire sky, and the number of dipoles was
sufficiently small that a simple software correlator could be used to
make all-sky images.  From~2006 October to~2007 February, we
conducted an all-sky transient search program, acquiring a total
of~106~hr of data; the time sampling varied, being 5 minutes at the
start of the program and improving to~2~minutes by the end of the
program.
We were able to detect solar flares, and in a special-purpose
mode, radio reflections from ionized meteor trails during the 2006
Leonid meteor shower.  We detected no transients originating outside
of the solar system above a flux density limit of~500~Jy, equivalent
to a limit of no more than about $10^{-2}$
events~yr${}^{-1}$~deg${}^{-2}$, having a pulse energy density $\ga
1.5 \times 10^{-20}$~J~m${}^{^-2}$~Hz${}^{-1}$ at~73.8~MHz for pulse
widths of about~300~s.  This event rate is comparable to that
determined from previous all-sky transient searches, but at a lower
frequency than most previous all-sky searches.  We believe that the
LWDA illustrates how an all-sky imaging mode could be a useful
operational model for low-frequency instruments such as the Low
Frequency Array, the Long Wavelength Array station, 
the low-frequency component of the Square
Kilometre Array, and potentially the Lunar Radio Array.
\end{abstract}


\section{Introduction}\label{sec:intro}

Transient emissions---in the form of bursts, flares, and pulses from
compact sources---are the signposts for explosive or dynamic events.
As such, transient sources offer insight into a variety of fundamental
aspects of physics and astronomy, ranging from studying the mechanisms
of particle acceleration on the \objectname{Sun} and nearby stars to tracking
stellar evolution and death across the Universe to probing the
intervening medium(a).

At radio wavelengths, there are well-known classes of transients, such
as the \objectname{Sun} and radio pulsars, as well as a long history
of observing transients from triggers at other wavelengths, such as
$\gamma$-ray burst (GRB) afterglows
\citep[e.g.,][]{pgtbfk97,tfksff98,frailetal00}, or the monitoring of
known transient sources such as X-ray binaries and microquasars
\citep[e.g.,][]{wg98,k-wfpbmmv02,mmrs02,rdm02}.  Further, a series of
observations and discoveries over the past decade have emphasized that
the radio sky may be quite dynamic.  Known sources have been
discovered to behave in new ways and what may be entirely new classes
of sources have been discovered---pulsed radio emission has been
observed from brown dwarfs \citep{hbl+07} and formerly radio-quiet
magnetars \citep{crh+06}; single or highly intermittent pulses have
been detected from neutron stars, also known as rotating radio
transients \citep[RRATs,][]{mll+06,kle+10,b-sb10}; intense giant
pulses have been detected from the \objectname{Crab} pulsar
\citep{hkwe03}; and several as-yet unidentified radio transients have
been found \citep{hlkrmy-z05,bsb+07,lbmnc07}.

There are also numerous classes of objects that, by extension of known
physics, have been hypothesized to be radio emitting, and which might
appear as radio transients.  These include extrasolar planets
\citep{fdz99}, magnetar flares or giant pulses from pulsars in other
galaxies \citep[e.g.,][]{mc03}, prompt emission from GRBs
\citep{uk00,sw02}, evaporating black holes \citep{r77}, and
extraterrestrial transmitters \citep{cls97}.

Meter-wavelength observations have a long history of being used for
studying the time domain.  For example, solar radio emissions have
long been known to be dynamic, and pulsars were first discovered in
81~MHz observations \citep{hbpsc68}.
Advantages of low-frequency instruments for surveying the transient
sky include naturally large instantaneous fields of view and
sensitivity to steep spectral index sources such as might result from
coherent emission processes, while disadvantages include radio-wave
propagation effects such as dispersion and multi-path propagation,
absorption either within the source or along the line of sight, 
ionospheric disturbances, and radio frequency interference (RFI).
A number of new telescopes with meter-wavelength
capabilities are under construction, including the Murchison
Wide-field Array \citep[\hbox{MWA},~][]{lcm+09}, the \anchor{http://www.lofar.org}{Low Frequency Array (LOFAR)}, the Precision
Array to Probe the Epoch of Reionization
\citep[\hbox{PAPER},~][]{pbf+10}, and the Long
Wavelength Array \citep[\hbox{LWA},~][]{ellingson09}---many of the
science cases for these telescopes explicitly include searches for
radio wavelength transients.
It is anticipated that the success of these
telescopes will motivate the low-frequency component of the Square
Kilometre Array \citep[\hbox{SKA},~][]{dhsl09}
 and the Lunar Radio Array \citep[\hbox{LRA},~][]{lchfb09}.

This paper presents an all-sky monitoring campaign for radio
wavelength transients conducted on the Long Wavelength Demonstrator
Array (LWDA).  The objective is two-fold, both to detect or
constrain radio transients from the observations and to illustrate how
future instruments could be used for all-sky searching and
monitoring observations.  The plan of this paper is as follows.  In
\S\ref{sec:lwda} we describe the LWDA itself and its data acquisition
path, in \S\ref{sec:all-sky} we describe the LWDA data acquisition and
imaging pipeline specific to the formation of essentially all-sky
images and radio transient searching, in \S\ref{sec:results} we
present our results and analysis, in \S\ref{sec:future} we discuss
possibilities for future instruments, and in \S\ref{sec:conclude} we
summarize our conclusions.

\section{The Long Wavelength Demonstrator Array}\label{sec:lwda}

\begin{figure*}[bth]
\begin{center}
\includegraphics[width=0.9\textwidth]{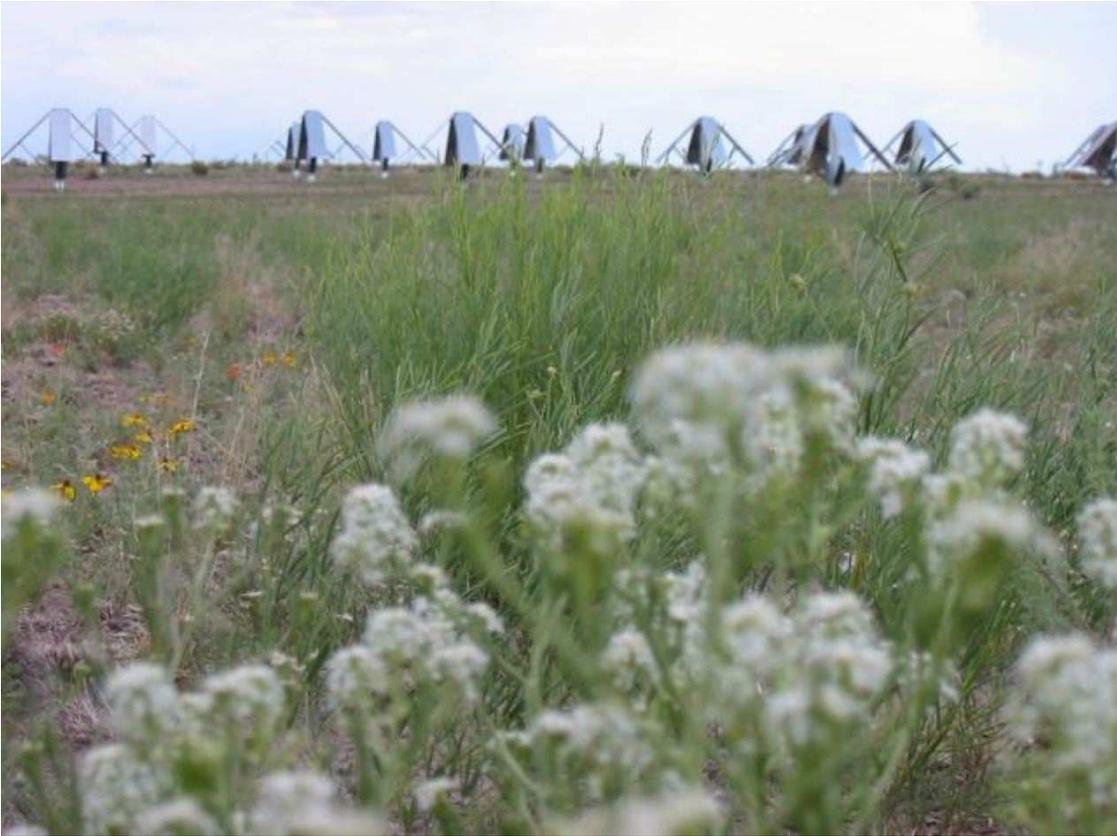}
\end{center}
\vspace*{-5ex}
\caption{Photo of the Long Wavelength Demonstrator Array in New Mexico
  showing the individual dual-polarization stands. Each stand consists
  of four droopy blades which provide dual-polarization capabilities.}
\label{fig:LWDA}
\end{figure*}

\begin{figure}[tbh]
\begin{center}
\includegraphics[angle=-90,width=0.9\columnwidth]{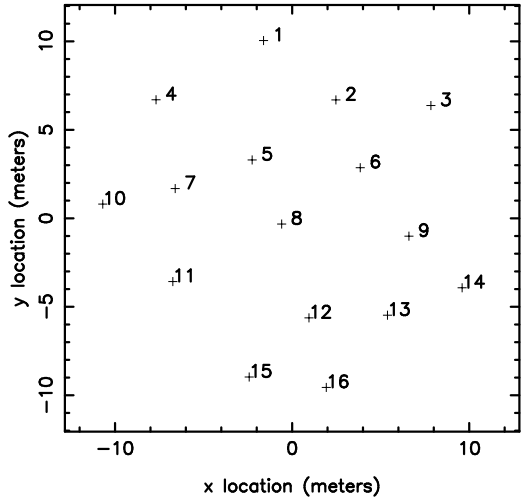}
\end{center}
\vspace*{-3ex}
\caption{Location of the antennas within the \hbox{LWDA}.}
\label{fig:array}
\end{figure}

The Long Wavelength Demonstrator Array (LWDA) was a testing platform
developed by the Naval Research Laboratory prior to construction of
the \hbox{LWA}.
The LWDA site work included fielding prototype hardware for the
\hbox{LWA}, development of site preparation techniques, continual
monitoring of the RFI environment, software development, and initial
science.
Located on the Plains of San Agustin in New Mexico near the center of
the National Radio Astronomy Observatory's Very Large Array (VLA), the
LWDA consisted of~16 dual-polarization dipole \emph{stands} operating as a
phased dipole array with a frequency range of~60--80~MHz
\citep[Figure~\ref{fig:LWDA};~]{y+07}.  The element locations represented
a compact subset of a larger 256 stand pseudo-random distribution that was
developed for the LWA \citep{kc09}.  The maximum baseline for the LWDA
was approximately 20~m (Figure~\ref{fig:array}), providing an angular
resolution at the zenith of about~12\arcdeg\ at~74~MHz.

\begin{figure}[tb]
\plotone{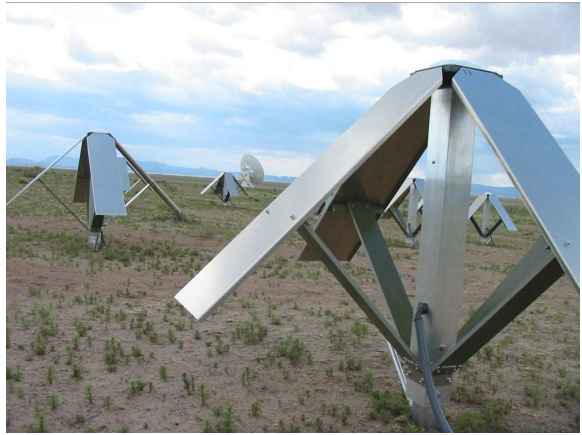}
\vspace*{-5ex}
\caption{Photo of several LWDA elements showing the droopy blades and
  an NRAO VLA antenna in the background.}
\label{fig:blades}
\end{figure}

Each of the 16 LWDA dipole stands consisted of~2 ``droopy,'' fat
dipoles, one per polarization, with the dipoles mounted at a
45\arcdeg\ droop angle to a central post (Figure~\ref{fig:blades}).
Each dipole consisted of two ``blades,'' each 28~cm wide by~1.05~m long.
This design was chosen to increase the symmetry in the $E$- and
$H$-planes \citep{ke_18}, broaden the antenna beam pattern (i.e.,
increase antenna sky coverage), and obtain a larger operational
bandwidth.

The radio frequency (RF) signals from each dipole were amplified by an
active balun (24~dB of fixed gain) before being passed to the LWDA
electronics, located within a central, shielded shelter.  Inside the
shelter, the incoming RF data stream entered the digital receiver
signal processing chain.  The signals were digitized by a
dual-channel 10-bit analog-to-digital converter (ADC)
at~100 megasamples per second.

The LWDA electronics had the capability to provide digital delay
beamforming of two fully independent beams of bandwidth 1.6~MHz each
as well as the capability for all-sky monitoring. This capability was
implemented through two identical signal processing chains that had
independent control parameters to allow for two frequencies, two
spatial beams, or two polarizations. The first stage of digital
processing was a first-in--first-out (FIFO) pipe that provided an
integer sample delay from~0~ns to~150~ns in steps of~10~ns.  This
delay provided partial compensation of geometric path and system time
delays at this stage.  Following the FIFO was a complex mixer where the
in-phase (I) and quadrature (Q) components for each signal chain were
generated using a CORDIC rotator.  The signals then passed through a
low-pass filtering stage using a Cascaded Integrator Comb (CIC) filter
and were decimated by a factor of~14.  Next, the signals passed through
another \hbox{FIFO}, which provided coarse sample delays in excess
of~2~$\mu$s in~140~ns increments.  The final 1.6~MHz bandpass for each
data stream was set in a finite impulse response (FIR) filter that also
reduced the data rate by another factor of~3. Following this stage,
the data were interleaved and passed to an adder board at the final
data rate of approximately 4.8~\hbox{MSPS}. The adder board was used to
either sum (beamform) or interleave (all-sky image) the incoming
coherent signals.

The LWDA frequency range was chosen, in large part, because early plans
for the array included the goal of operating it as one of the elements
of the VLA 74~MHz system \citep{kle+07}, which in turn was centered
on the (primary) frequency allocation for radio astronomy
at~73.0--74.6~MHz in the U.{}S.  The operational frequency range was
much broader and includes frequency allocations for a number of
different services, primarily TV broadcasting, at the time of the LWDA
operation.\footnote{
Since the time of these experiments, the U.{}S. has converted from
analog to digital TV broadcasts, and much of the LWDA operating
frequency range is no longer used for TV broadcasts at the time of
writing \citep{patcrane09}.
} Tests with the LWDA and the prototype LWA equipment have shown the
presence of TV signals, but few other strong emissions in this band
\citep{dr07,hrpde07,phre07,ryk+07,c08,jm??}.  Further, experience with
the VLA 74~MHz system indicated that, for that system, the most
significant source of interference was the VLA itself.  Indeed, in
comparison with the \hbox{VLA}, for which the 74~MHz feeds were mounted
near prime focus (several meters above the ground), the LWDA should be
less susceptible to \hbox{RFI}.  Our experience with the LWDA was
consistent with the experience from the \hbox{VLA}, in that no
significant amount of observing time was lost due to \hbox{RFI}. (See
also \S\ref{sec:detect}.)

The LWDA system was sky-noise dominated by at least 6~dB over its full
operating range.
The LWDA construction was completed and first-light images were
obtained on~2006 October~23.

Our focus here is on the all-sky observations obtained by the
\hbox{LWDA}, but it also had a  two-element interferometer mode.  In
this mode, an outlier dipole could be utilized  to monitor the flux densities of isolated point sources,
contributing to a study of the secular flux density decrease of
\objectname{Cas~A} \citep{hk09}.

\section{All-Sky Observations}\label{sec:all-sky}

The observations presented here consist of operating the LWDA in its
``all-sky'' imaging mode observing at zenith.  In this operational
mode, the all-sky data acquisition system cycles serially through all
120 pairs of dipoles in the array and correlates the data from each
pair.  Each baseline is measured for~51~ms with a total cycle time
of~13~s for both polarizations of all baselines (including
auto-correlations).  As a result of various processing overheads, the
on-sky cadence of imaging described herein is slower.  Our
observations began with a 5~minute cadence (i.e.,
13~s to acquire the full set of cross correlations to form an
image with images repeated every 5~minutes), which was later improved
to a 2~minute cadence for most of the data reported here.

Our focus on an (all-sky) \emph{imaging} transient search stems from
the fact that the number of dipoles in the LWDA was relatively small
so that cross-correlation of all of them was computationally tractable.
The alternative to an imaging search, in which transients are
identified by comparing images of the sky at different epochs, is a
\emph{non-imaging} search, in which transients are identified in time
series of the voltages or intensities from a telescope(s).
Traditionally, the tradeoffs between the two kinds of searches have
been that imaging searches obtain higher angular resolution at the
cost of time resolution with the converse being true for non-imaging
searches.  The choice between the kind of transient search to employ
depends not only on the instrument available, but also upon what is
known about the transient population(s) of interest \citep{n03}.

The observations were centered at a frequency of~73.8~MHz with~80
spectral sub-bands over a total bandwidth of~1.6~MHz (20~kHz per
sub-band).  During the course of the array installation, the amplitude
and phase of the signal path from each dipole were determined.  These
values were applied in real time as the cross-correlations
(visibilities) were formed.  The normalized bandpass shapes were
applied to correct for individual telescope variations.  The
visibilities were then recorded on disk for later processing.

\begin{figure*}
\begin{center}
\includegraphics[width=0.95\textwidth]{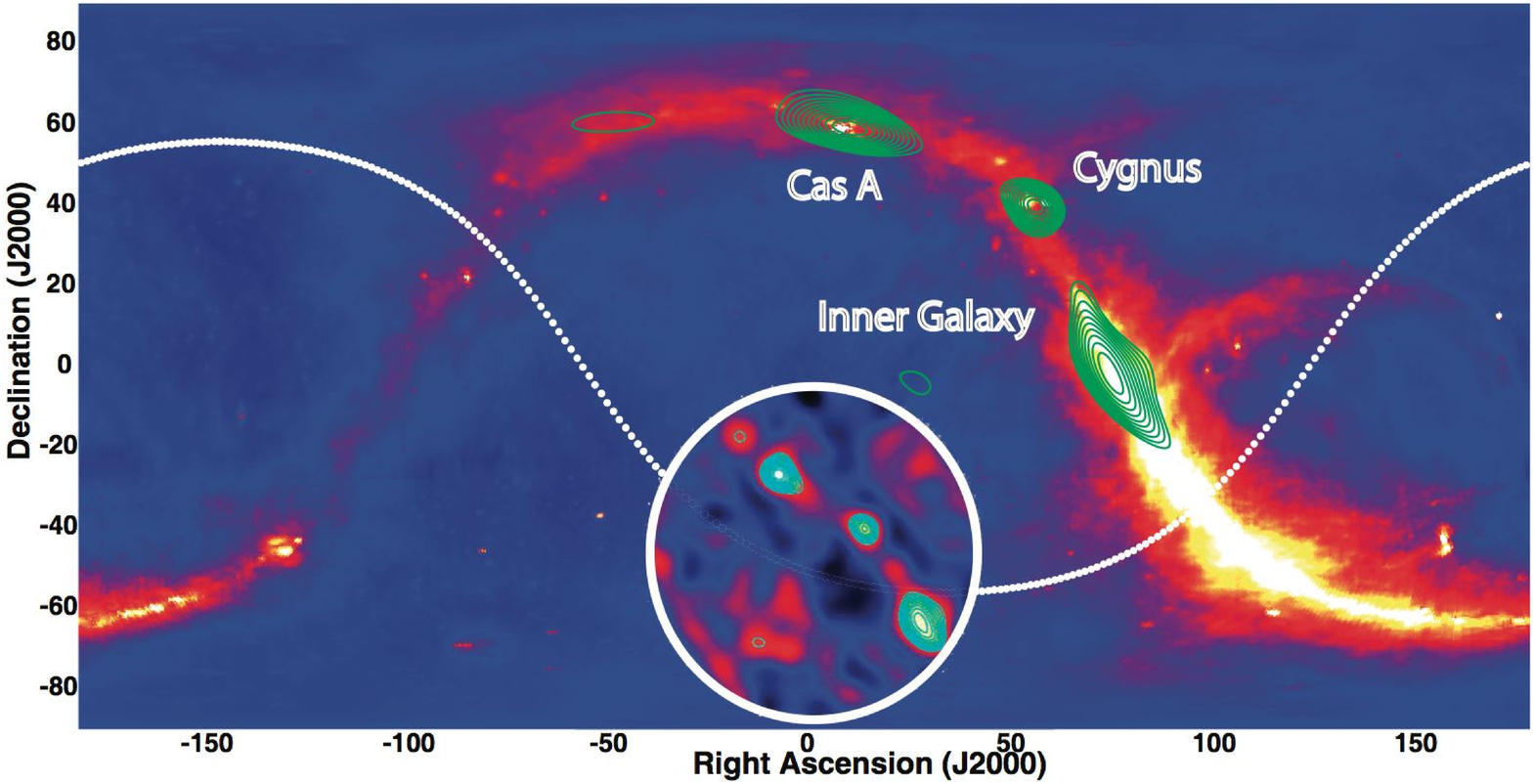}\\
\includegraphics[width=0.95\textwidth]{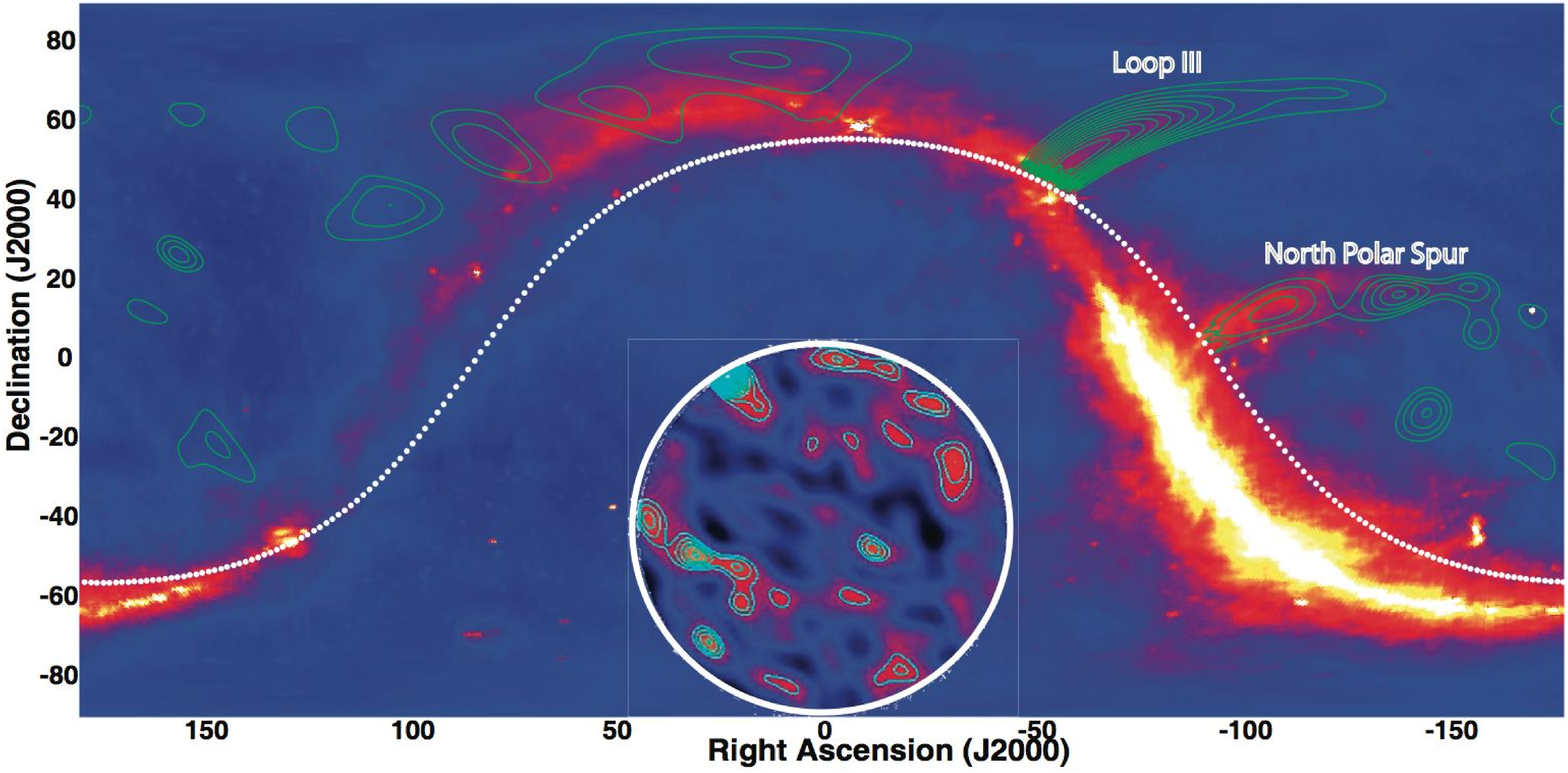}
\vspace*{-5ex}
\end{center}
\caption{Comparison of all-sky LWDA images (green contours) with the
408~MHz all-sky image (color) from \cite{hssw82}.  Insets in both
panels (demarcated by the white circle) show the all-sky LWDA images
(in color); in these insets, north to up and east is to the left.  The dotted white line indicates the horizon for the \hbox{LWDA}, with the portion above the dotted white line being visible.  The LWDA images are at~73.8~MHz and were acquired on~2006
October~28.  
{Top}: a snapshot acquired when the Galactic plane was passing
nearly overhead.  Individual, well known sources and the inner
Galactic plane dominate the image.
{Bottom}: a snapshot acquired when the inner Galactic plane
was below the horizon.  While the LWDA image would appear to be only
noise, there are features that can be identified with the
\protect\objectname{North Polar Spur} and \protect\objectname{Loop~III}
radio structures.
}
\label{fig:all-sky}
\end{figure*}

Post-processing consisted of the following steps:
\begin{enumerate}
\item The raw visibilities were converted into the FITS-IDI format
	\citep{cllw07}.
\item The visibility files were ingested by the Astronomical Image
	Processing System (AIPS).\footnote{
version 31DEC08}
\item Standard coplanar imaging procedures were used to convert
	the visibilities to images.
\item Strong sources in the images were blanked (see below), and, because the
	images formally extend beyond the horizon, they were clipped
	at the horizon.
\end{enumerate}
Figure~\ref{fig:all-sky} presents a typical all-sky image.  Sources
easily detected in the LWDA all-sky maps include \objectname[]{Cas~A},
\objectname[]{Cyg~A}, the inner Galactic plane, \objectname[]{North
Polar Spur}, \objectname[]{Loop~III}, and the \objectname[]{Sun}.

We used the apparent brightness of both \objectname[]{Cas~A} and
\objectname[]{Cyg~A} to obtain a crude estimate of the power pattern
of the \hbox{LWDA}.  Simulations of the power pattern of the
individual dipoles \citep{y+07} suggest that the gain decreases toward
the horizon (as would be expected from a dipole) as well as toward the
zenith.  The latter is caused by the design goal of operating the
dipoles over a relatively large frequency range.  The brightness
measurements of \objectname[]{Cas~A} and \objectname[]{Cyg~A} do show
these large-scale gain variations as well as smaller variations, which
we estimate to be approximately a factor of~2 (3~dB) in amplitude.

The all-sky transient data were passed into an analysis pipeline
written within the AIPS software system. The pipeline first clipped
high phase-center amplitudes that are known to be generated by
sporodic monitor and control (M\&C) software issues. The data were then
Fourier transformed into dirty images of the entire sky visible at the
LWDA site.  Even though the LWDA can observe essentially the entire
sky, ``wide-field'' imaging techniques \citep{cp92} are not required
given the size of the LWDA, its operational frequency, and our use of
snapshot imaging.

Images were blanked around the positions of the strong sources that
could be detected easily (i.e., \objectname[]{Cyg~A}, \objectname[]{Cas~A}, the \objectname[]{Sun}, the inner Galactic plane, \objectname[]{Loop~III}, and
the \objectname[]{North Polar Spur}).  This strategy of blanking the
images allowed us to use a relatively simple threshold test to search
for the presence of transients, at the cost of reducing somewhat our
sky coverage.  Near the end of the LWDA's operation, we did attempt to
change the array so that the time sampling was in sidereal time.  Had
this been successful, it would then have been possible to subtract
images acquired on different days and construct ``difference'' images
for which this blanking scheme probably would not have been necessary.

As a measure of our ability to detect a transient, particularly in
light of our non-simultaneous sampling of the baselines, we conducted
a number of simulations in which transients of different strengths and
positions within the image were inserted into the data.
Table~\ref{tab:simulate} summarizes the results: Column~1 indicates
what fraction of the baselines are illuminated by the simulated
transient, Column~2 shows the implied duration of the transient, and
Column~3 shows the approximate signal-to-noise ratio (S/N) that the
transient would have to have in order to be detected.  One important
aspect of these simulations is that the simulated transients were
inserted into the visibility data.  As such, these simulations do not
take into account the antenna power pattern.  The trend in
Table~\ref{tab:simulate} of increasing S/N required in order to detect
the transient is perhaps not surprising, but we find it reassuring
that we could detect transients, even those whose durations are only a
minor fraction of the time it took the LWDA to acquire a full sample
of visibility data.

\begin{deluxetable}{lcr}
\tablewidth{0pc}
\tablecaption{Transient Detection Simulations\label{tab:simulate}}
\tablehead{%
 \colhead{Baseline Fraction} & \colhead{Time} & \colhead{S/N} \\
 \colhead{Illuminated}       & \colhead{(s)}  & }
\startdata
100\%  &           6      &          5 \\
 90\%  &           5.4    &          7 \\
 75\%  &           4.5    &         10 \\
 50\%  &           3      &         13 \\
 30\%  &           1.8    &         30 \\
\enddata
\tablecomments{The results of injecting simulated transients into the
LWDA data.  Column~1 indicates what fraction of the baselines are
illuminated by the simulated transient.  Because the LWDA acquired
all-sky imaging data by sampling baselines sequentially, Column~2
shows the implied duration of the transient.  Column~3 shows the
approximate signal-to-noise ratio (S/N) that the transient would have
to have in order to be detected.}
\end{deluxetable}

Using our simulations and visual inspection of the images as a guide,
each all-sky snapshot image was searched for events above the
5$\sigma$ level.  Any image containing a potential transient event was
saved for further analysis.

\section{Results}\label{sec:results}

We begin this section by both motivating and illustrating the
capability of the LWDA to detect radio transients, and we then turn to
the detection of astronomical radio transients.

\subsection{Detected Radio Transients: The Sun and Meteor
	Trails}\label{sec:detect}

\begin{figure*}
\begin{center}
\includegraphics[width=0.47\textwidth]{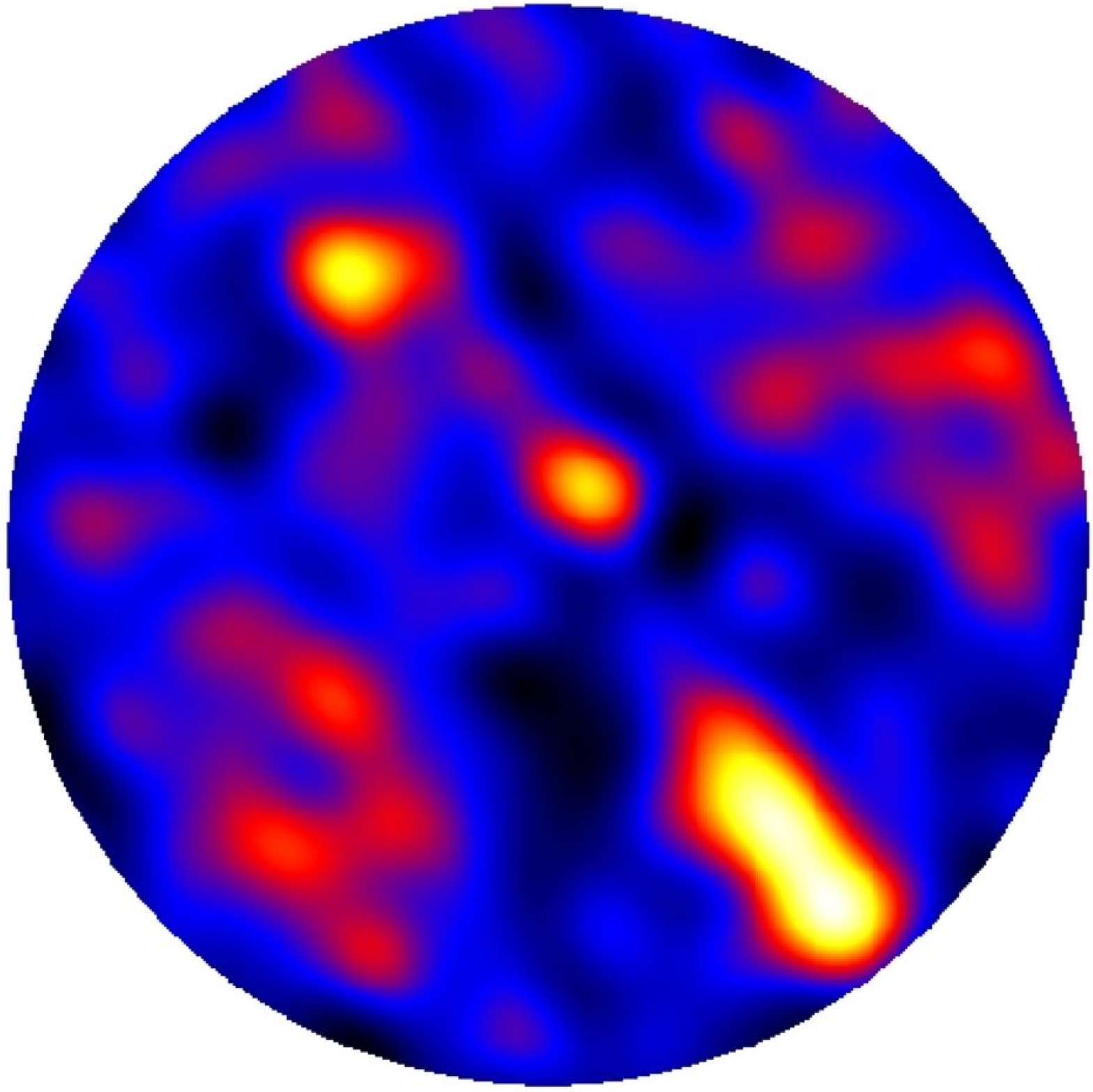}\,\includegraphics[width=0.47\textwidth]{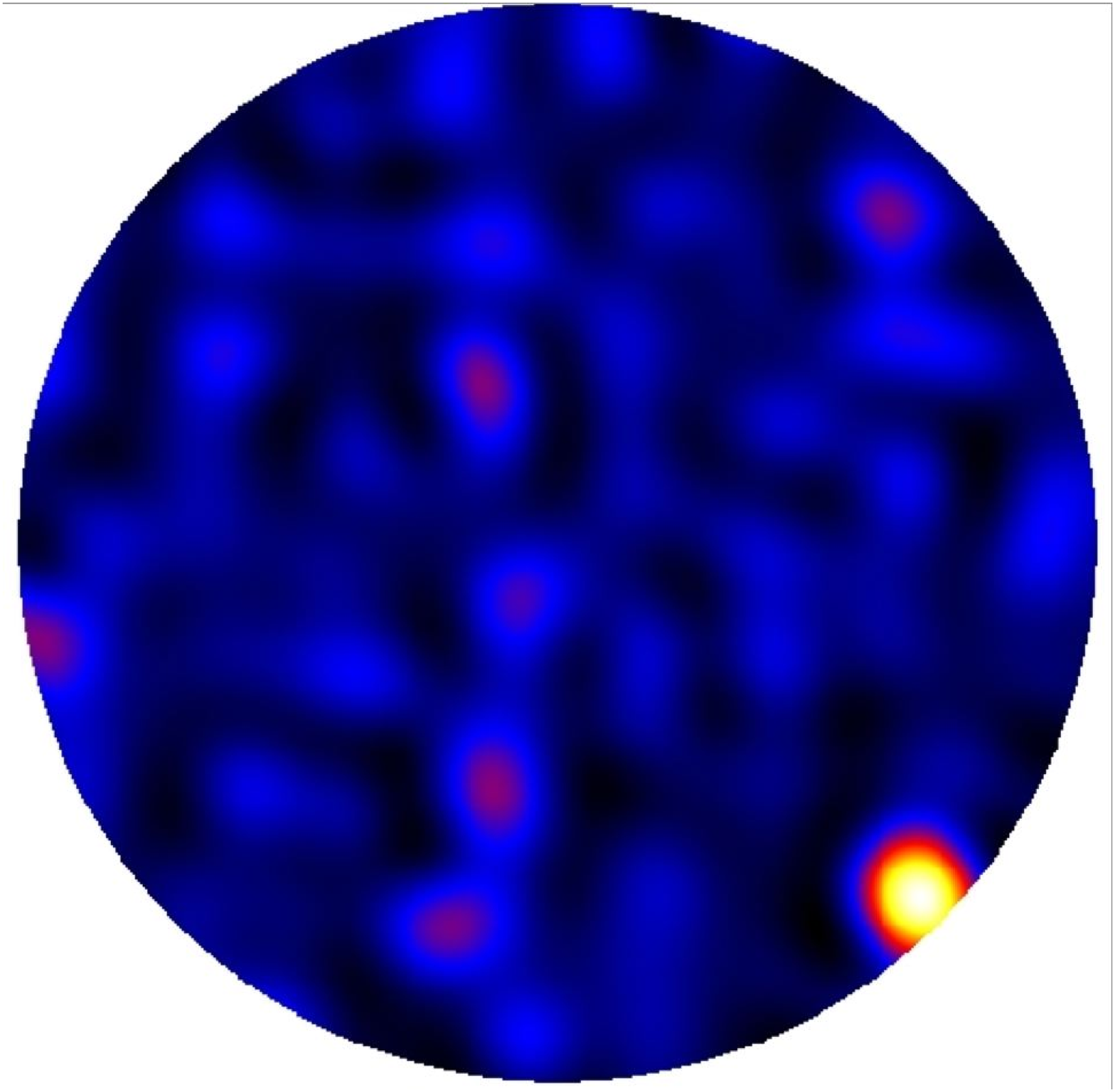}
\end{center}
\vspace*{-5ex}
\caption{Illustration of the LWDA's detection of a solar radio burst.
{Left}: the sky before the solar radio burst.
{Right}: the sky during the solar radio burst of~2006
December~14.  This example is one of the several intense radio bursts that
occurred in~2006 December.}
\label{fig:sun}
\end{figure*}

The \objectname{Sun} is a well-known variable radio source at tens of
MHz \citep[e.g.,][]{wm50}.  While the
\objectname{Sun} is currently in a state of exceptionally low radio
emission, during the course of the LWDA observations reported here, a
number of solar radio bursts were detected, including a series of
intense radio bursts on~2006 December~14 (Figure~\ref{fig:sun}),
probably associated with the X-class flares that occurred then.

The focus of our monitoring program was primarily on possible
extrasolar system transients, and, as such, we made no effort to track
the solar flux density (\S\ref{sec:all-sky}).
However, a similar monitoring program could be used to track the flux
density of the Sun at these frequencies.  Such a program would be more
valuable if the array monitoring the \objectname{Sun} had both a larger
bandwidth and a higher time resolution.  With these improved
characteristics, not only would such an array be capable of detecting
solar radio bursts, it might be able to image their evolution in frequency.

Conversely, as Figure~\ref{fig:sun} illustrates, the \objectname{Sun}
can become a source of ``interference.''  During these intense solar
radio bursts, the dynamic range of the LWDA and imaging pipeline were
sufficiently limited that only the \objectname{Sun} was visible.
Whether other observations can be conducted during solar radio bursts
will depend upon the imaging dynamic range of future systems, however,
exceeding an imaging dynamic range of order 100 seems difficult for
rapid, all-sky imaging pipelines.

Another well-known class of radio transients is reflections from
ionized meteor trails.  As a meteor enters Earth's atmosphere, its
velocity is high enough that it can produce an ionized trail, which in
turn can reflect a radio transmission \citep[e.g.,][]{mmb48}.  

\begin{figure*}
\begin{center}
\includegraphics[width=0.47\textwidth]{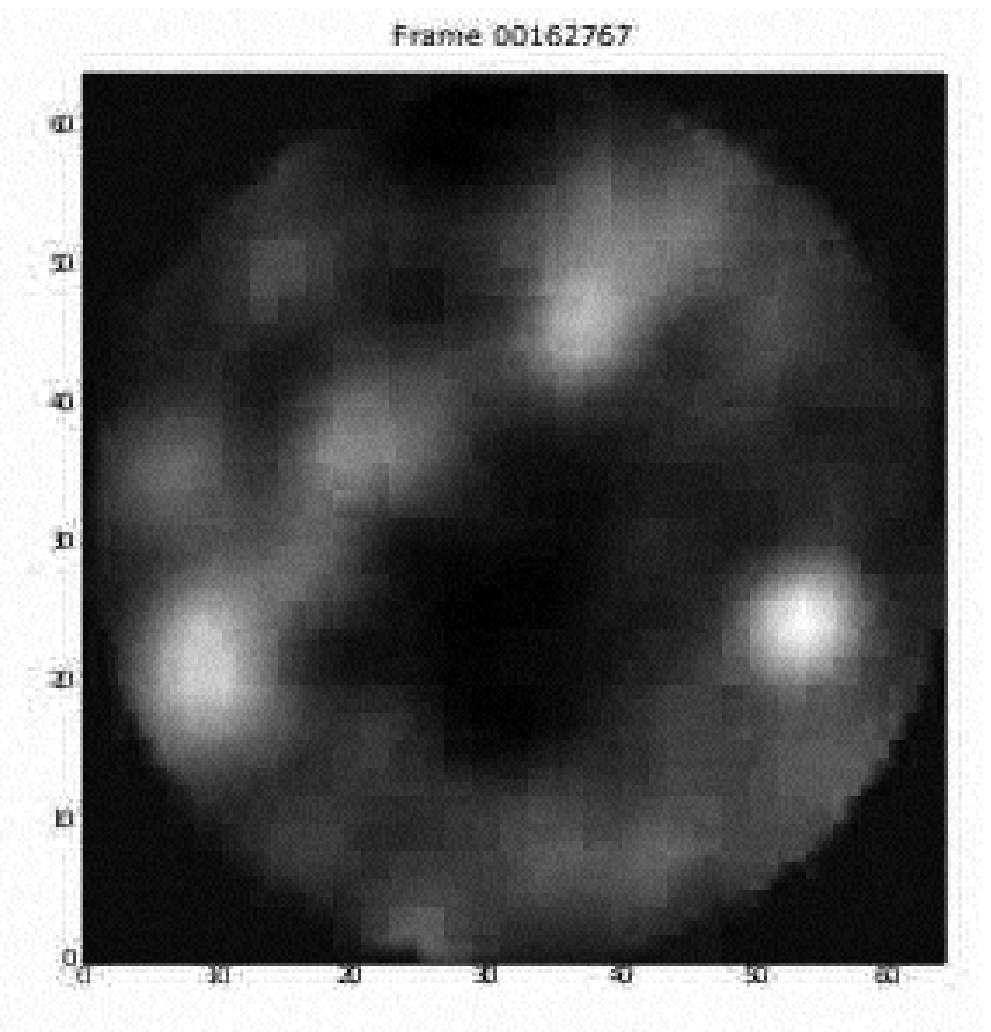}\,\includegraphics[width=0.47\textwidth]{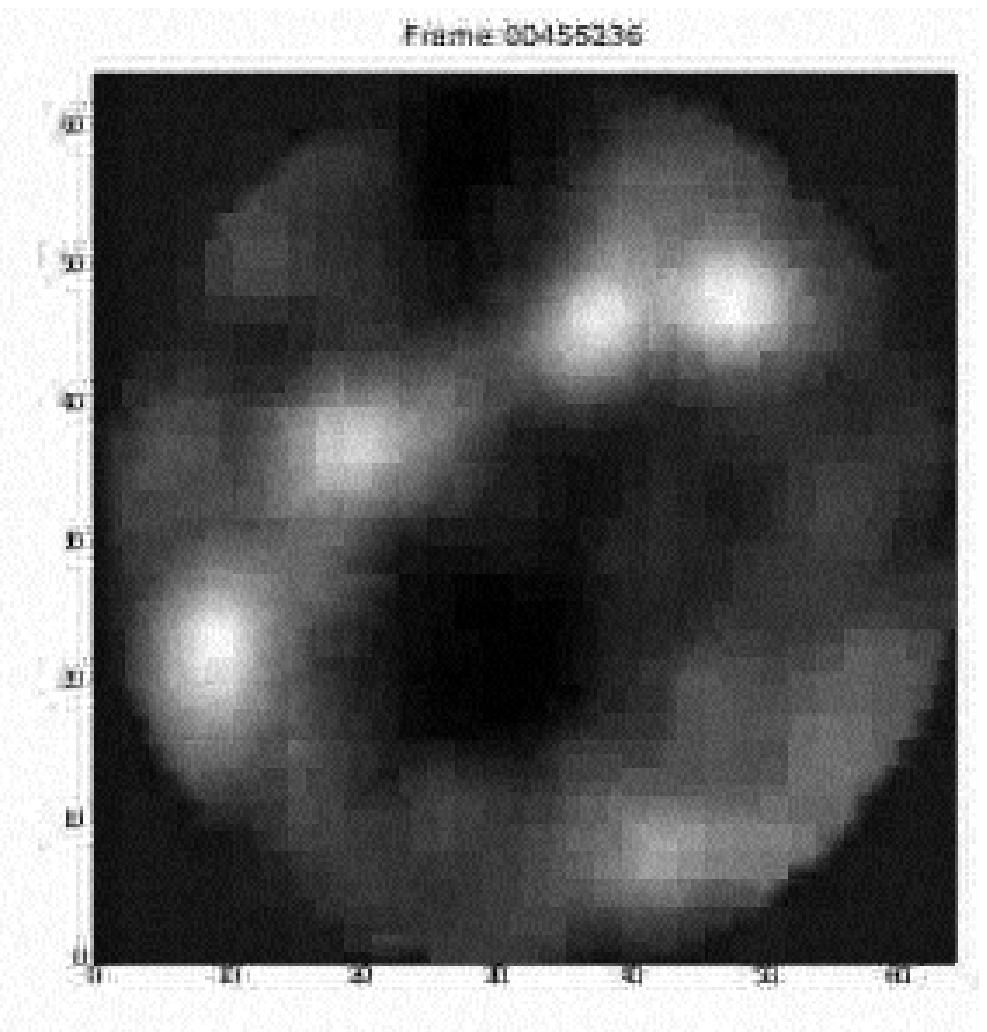}\\
\includegraphics[width=0.6\textwidth]{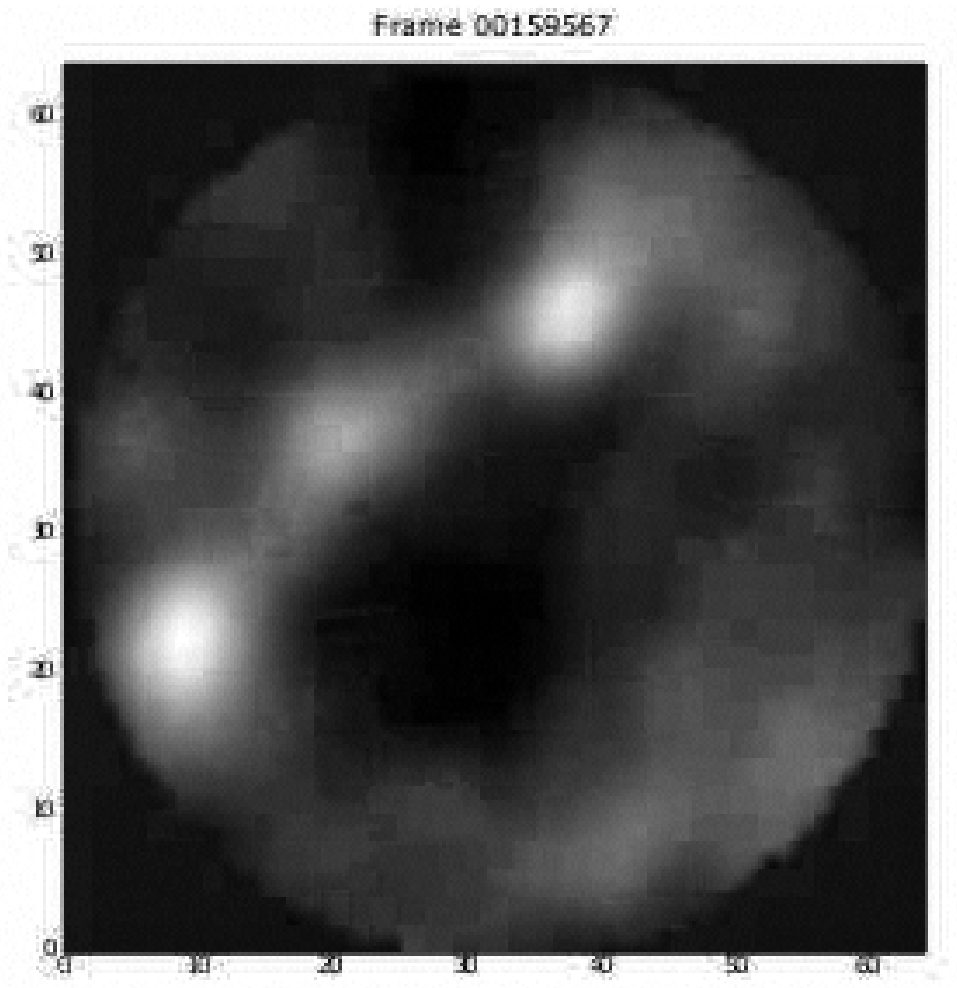}
\end{center}
\vspace*{-5ex}
\caption{%
LWDA images of reflected TV signals from ionized meteor
trails during the 2006 Leonid meteor shower.  Each image is
constructed from a 0.1~s data acquisition at a frequency of~61~MHz.
{Top left} and {top right}: images of reflected
signals from Leonid meteors.
{Bottom}: a reference image showing the nominal sky observed
by the \hbox{LWDA}.  The zenith is at the center of the image.  The
two discrete emitters near the center of the field are celestial radio
sources (\protect\objectname{Cas~A} and \protect\objectname{Cyg~A}).  The band
stretching from the upper right to lower left, terminating in the
strong emission near the edge of the field is from the Milky Way.}
\label{fig:meteor}
\end{figure*}

In~2006 November, during the Leonid meteor shower, data were acquired
with the LWDA tuned to a frequency of~61~MHz.  At the time of these
observations, this frequency was within the band allocated for
television broadcasting, and there were a number of TV stations in New
Mexico and the southwest U.{}S.\ that used this allocation.  It is a
simple matter to show that even a TV transmitter of relatively modest
power located more than 1000~km distant would still produce a strong
reflection.  For the purposes of this observation, the process for
data acquisition was modified to reduce the number of bits and allow
simultaneous sampling of all 16 dipoles.  Figure~\ref{fig:meteor}
illustrates two examples of a meteor reflection imaged by the
\hbox{LWDA}.

\subsection{Astronomical Radio Transients}\label{sec:astro}

The observing campaign lasted from~2006 October~27 to~2007
February~17, during which there were 59~days on which some useful data
were acquired.  On~2006 December~13, we were able to improve the
cadence at which images were formed from~5~minutes to~2~minutes.  The
resulting total observing time for the transient campaign was
106~hr.  During these observations, no transients were detected
above the 5$\sigma$ level in any image.  Based on comparative analysis
of the brightnesses of
\objectname{Cas~A} and the \objectname{Sun}, we estimate the
(1$\sigma$) noise level
in the LWDA all-sky snapshots to be roughly 500~Jy.
We also assume that the effective sky coverage of the LWDA is roughly
10,000~deg${}^2$ ($\sim \pi$~sr).  Although, we show essentially
images of the entire hemisphere, the gain of the LWDA was unlikely to
be constant across the sky, an assumption partially supported by
modeling.  Consequently, we assume that the effective sky coverage is
somewhat less than a full hemisphere.

A total of 29,437 data sets were processed through the imaging
pipeline.  Of these, 1764 produced potential transient candidates.
Visual inspection of these showed that they could be explained
as non-astronomical transients:  slight errors in our
calculations of the extents of strong sources resulted in one
pixel of a strong source not being blanked appropriately.

Combining these results, we place a limit of $\la 10^{-2}$
events~yr${}^{-1}$~deg${}^{-2}$, having a pulse energy density $\ga
1.5 \times 10^{-20}$~J~m${}^{-^2}$~Hz${}^{-1}$ at~73.8~MHz for pulse
widths of about 300~s.  We now consider multiple characterizations of
this result for easy comparison with existing results in the
literature.  Our focus is on comparison with MOTOR \citep{alv89} and
STARE \citep{khcm03}, as both represent nearly all-sky surveys at
frequencies below~1~GHz.

\begin{figure}[t]
\begin{center}
\includegraphics[angle=-90,width=0.95\columnwidth]{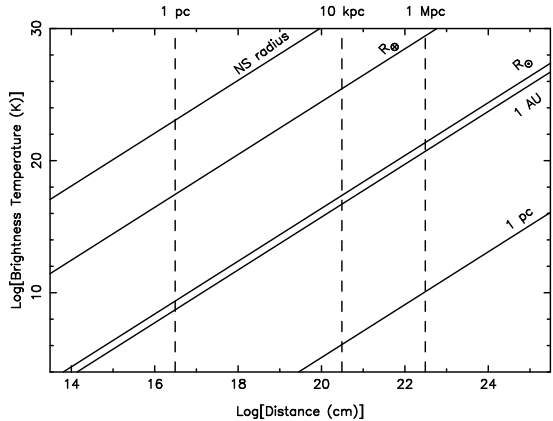}
\end{center}
\vspace*{-3ex}
\caption{Brightness temperature limits for objects of varying
sizes (diagonal lines) as a function of the distance to the object.
Lines for fiducial object dimensions are shown as are fiducial
distances.  The required brightness temperature for an object to be
detectable is above and to the left of a line of constant object
dimension.  The range of the ordinate is approximately that allowed
for the LWDA observations (viz.\ Figure~\ref{fig:phasespace}).}
\label{fig:transients2}
\end{figure}

Following \cite{khcm03}, the LWDA flux density limit can be recast in
terms of the brightness temperature that an object would have to have
in order to be detectable.  Using parameters appropriate for the
\hbox{LWDA}, and assuming a quite conservative 5000~Jy flux density
limit (10$\sigma$), we find
\begin{equation}
T_B \gtrsim 40\,\mathrm{K}\,\left(\frac{D}{L}\right)^2,
\label{eqn:tb}
\end{equation}
for an object of linear size~$L$ located at a distance~$D$.
Figure~\ref{fig:transients2} shows the resulting brightness
temperature limits for transients detectable by the \hbox{LWDA}.
Sufficiently strong bursts from a neutron star or a ``super-flare''
from a nearby star are among the possible sources that might have been
detectable by the LWDA (or an instrument like the LWDA).  We consider
these possibilities further in \S\ref{sec:future}.

Table~\ref{tab:rate} compares our rate with those of other ``all-sky''
surveys at other wavelengths from the literature.  In constructing
Table~\ref{tab:rate}, we have restricted our focus to other, largely
similar search programs, namely those at frequencies below~1~GHz using
dipole-based arrays.  We have also restricted Table~\ref{tab:rate} to
programs that conducted blind surveys, as opposed to those that
targeted known sources (e.g., pulsars, GRBs).

\begin{deluxetable}{lcccccc}
\tablewidth{0pc}
\tabletypesize{\small}
\tablecaption{Transient Event Rates\label{tab:rate}}
\tablehead{%
                   & 
	& \colhead{Pulse}
	& 
	&                     & \colhead{Scaled} 
	& \\
 \colhead{Program} & \colhead{Rate Limit} 
	& \colhead{Energy Density}
	& \colhead{Timescale}
	& \colhead{Frequency} & \colhead{Pulse Energy Density} 
	& \colhead{Reference} \\
	           & \colhead{(events~yr${}^{-1}$~deg${}^{-2}$)}
	& \colhead{(J~m${}^{-^2}$~Hz${}^{-1}$)}
	& \colhead{(s)}
	& \colhead{(MHz)} & \colhead{(J~m${}^{-^2}$~Hz${}^{-1}$)}
}
\startdata
LWDA  & $10^{-2}$ & $1.5 \times 10^{-20}$ & 300   & 73.8 & \ldots 
	& this work \\
HM74  & 0.05      & $\sim 10^{-23}$       & $\sim 0.1$ & 270 & $3.4 \times 10^{-23}$ & 
	1 \\
K+77  & $4 \times 10^{-3}$ & $2.5 \times 10^{-22}$   & $> 1$  & 370--550 & $1.3 \times 10^{-21}$ & 
	2 \\
      & $4 \times 10^{-3}$ & $3 \times 10^{-22}$   & $> 1$  & 38--60 & $1.5 \times 10^{-22}$ & 
	2 \\
MOTOR & 163       & $10^{-28}$            & 0.025 & 843 & $1.1 \times 10^{-27}$
	& 3 \\
STARE & $10^{-3}$ & $10^{-24}$            & 0.125 & 611 & $8.3 \times 10^{-24}$
	& 4 \\
\enddata
\tablecomments{The \emph{scaled pulse energy density} is the 
pulse energy density scaled to a frequency of~73.8~MHz assuming a
nominal spectral index of~$-1$.  References: 
(1)~\cite{hm74};
(2)~\cite{kss+77};
(3)~\cite{alv89};
(4)~\cite{khcm03}.
}
\end{deluxetable}

In general, our LWDA search program produces an upper limit to the
event rate that is comparable to or competitive with other search
programs, typically at higher frequencies.  However, our serial data
acquisition scheme results in a somewhat poorer limit on the pulse
energy density.  If the LWDA data acquisition system had had a more
rapid sampling time, e.g., on a scale of order 0.1~s (the time scale
for the acquisition of a signal from an individual dipole), our pulse
energy density limit would have been comparable to many of the other
search programs.

\begin{figure*}
\begin{center}
\includegraphics[width=0.9\textwidth]{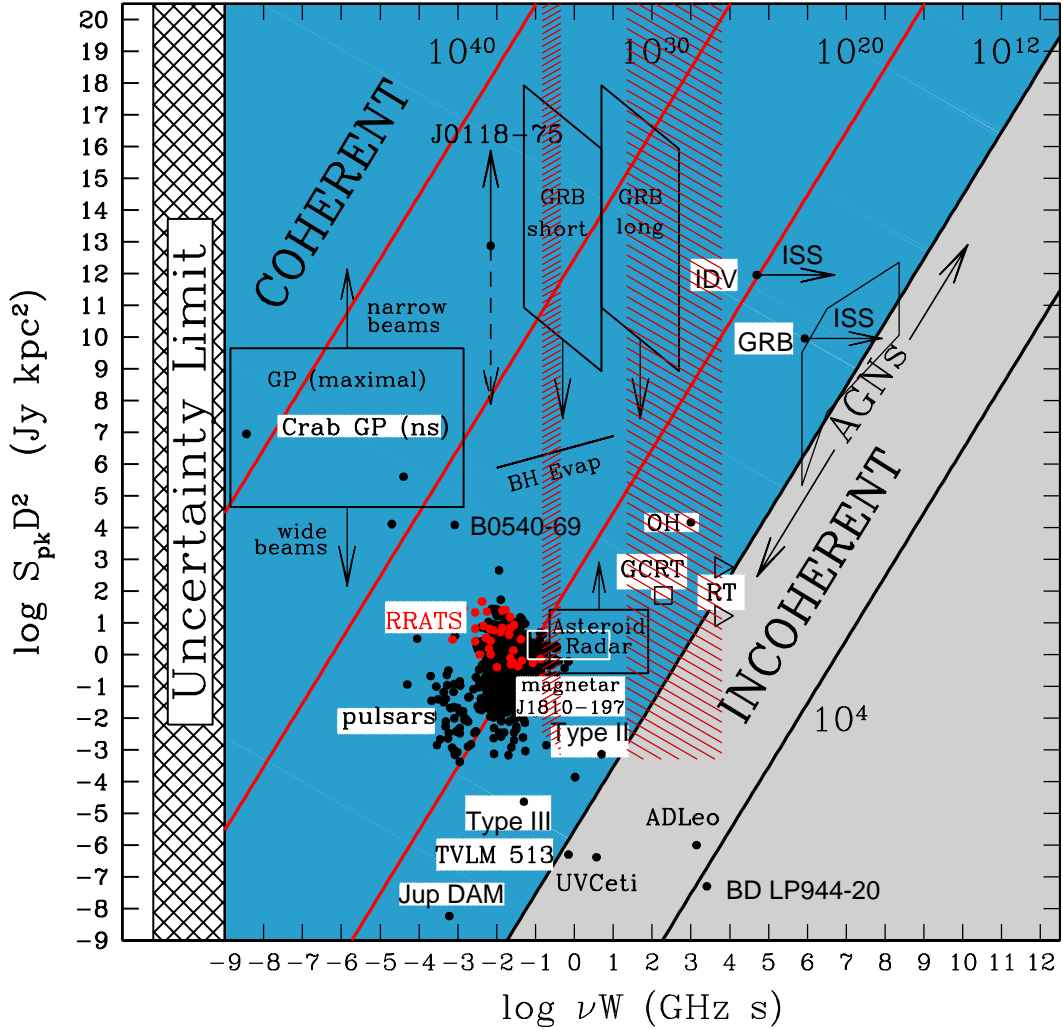}
\end{center}
\vspace*{-7ex}
\caption{%
Phase space defined by radio transients.  The abscissa is an
uncertainty-like quantity given by the product of the observing
frequency~$\nu$ and the duration of the transient~$W$.  For our LWDA
survey, we define two regions.  The first is from~2 to~6~s and is the
interval required to accumulate the visibility data from all of the
interferometric baselines (\S\ref{sec:all-sky} and
Table~\ref{tab:simulate}).  The second is from~5~minutes to~24~hr
and covers the cadence at which individual images were made to the
longest continuous observing duration during our program.  The
ordinate is a pseudo-luminosity~$S_{\mathrm{pk}}D^2$, for a peak flux
density~$S_{\mathrm{pk}}$ and distance~$D$.  We take a lower distance
limit of~1~pc, such as might be expected from a nearby low-mass star,
brown dwarf, or extrasolar planet.  Various classes of known or
hypothesized transients are indicated.  The sloping solid lines
indicate lines of constant brightness temperature, with a brightness
temperature of~$10^{12}$~K taken as the dividing line between coherent
and incoherent emission.}
\label{fig:phasespace}
\end{figure*}

An alternate means of characterizing radio transient searches is in
terms of a ``phase space'' diagram \citep{clm04}, which can be cast in
terms of the brightness temperature of a radio transient in the
Rayleigh-Jeans limit (Equation~\ref{eqn:tb}).
Figure~\ref{fig:phasespace} shows the region of phase space probed by
the LWDA in comparison with the range defined by various known and
hypothesized phenomena.  In producing this plot, we have assumed a
lower distance limit to any celestial radio transient to be 1~pc, such
as might be expected from a nearby low-mass star, brown dwarf, or
planet (\S\ref{sec:future}).  In practice, the local density of sources
may be sufficiently low that there are no transients within~1~pc,
which would have the effect of increasing the minimum value of $SD^2$
that the LWDA could have probed.
Clearly the upper distance limit could be
the edge of the observable Universe, which would imply correspondingly
higher brightness temperatures.

As a final approach to constraining the event rate for transients, we
consider a probabilistic one.  The typical determination of an event
rate (such as those in Table~\ref{tab:rate}) is derived from the total
observing duration~$T$ and the solid angle~$\Omega$ observed.
Instead, we ask what is the probability~$p$ of detecting a transient
in a given observation, given that no transients were detected in our
series of observations.

We conducted 29,437 trial observations during the course of this
observational program.  As
\cite{d86} discusses, the probability of obtaining $i$ detections
in~$N$ trials, given that the detection probability in any individual
trial is $p$, can be related to the beta distribution
\begin{equation}
P(n, j)
 = \frac{\Gamma(n+2)}{\Gamma(j+1)\Gamma(n-j+1)} p^j(1-p)^{n-j},
\label{eqn:prob}
\end{equation}
where $\Gamma(x)$ is the gamma function, $j \equiv \sum_{i=1}^n
X_i$, and $X_i = 1$ for a detection or~0 for a non-detection.  That
is, $j$ is simply the count of the total number of detections.  

With $j=0$, and integrating Equation~(\ref{eqn:prob}) to form the
incomplete beta function, we can determine the maximum detection
probability~$p$ for a specified confidence level.  For the LWDA or an
array with LWDA-like parameters ($\approx$ 10,000~deg${}^2$ sky
coverage, 73.8~MHz frequency, $< 300$~s duration), we find that the
probability of detecting a transient in a single LWDA observation
cannot exceed $p = 2 \times 10^{-4}$ lest the probability that we
should have detected a transient in the full LWDA transient campaign
exceeds 99.7\%.

Given the relatively large solid angle accessed by the \hbox{LWDA},
only a modest improvement in the probability of detecting transients
would result from increasing the sky coverage for future instruments
(or would have resulted from increasing the sky coverage of the LWDA
itself).  Rather, an increased probability is likely to result only
from improved sensitivity, developing a search method that does not
require blanking the Galactic plane, or both.

\section{Future Possibilities}\label{sec:future}

In the near future, the first station of the Long Wavelength Array
(LWA-1), the \hbox{LOFAR}, and the International LOFAR Telescope (ILT) will become
operational, with LOFAR being a subset of the \hbox{ILT}.  All will
consist of dipole stations having an operational frequency range
overlapping or comparable to that of the \hbox{LWDA}.

The individual dipole stations of LOFAR and the ILT will have a factor
of a few more dipoles than the LWDA while LWA-1 will 256 dipoles
(16$\times$ as many), but they may not be able to improve upon our
flux density limits substantially because the diameters of the
\hbox{LWDA}, LWA-1, and LOFAR stations are sufficiently small that
they are confusion limited on short time scales.  However, with
multiple LOFAR stations, the identification and excision of RFI may be
improved substantially \citep{bccl05}.
In the case of the LWA-1, all-sky imaging will be carried out by
passing a narrow bandwidth ($\sim 100$~kHz) signal from each of the 256 stands to a dedicated cluster for correlation and imaging.  This observing mode operates in tandem with the beamforming modes so that it will be possible to image the visible hemisphere full time.

The MWA will also
conduct transient searches, at a somewhat higher frequency range than
the \hbox{LWDA}, but it will likely not include an all-sky imaging
capability.  Building upon these next-generation low-frequency arrays
will be the \anchor{http://www.skatelescope.org}{SKA} and the \hbox{LRA}.  Their designs will be
influenced by the work on \hbox{LOFAR}, \hbox{MWA}, and similar low
radio frequency interferometers, but, in both cases, we anticipate
that transient searches, and potentially all-sky imaging, will be a
capability.  As Figure~\ref{fig:phasespace} illustrates, and as we now
discuss below, in principle, an imaging dipole station, similar to the
\hbox{LWDA}, could probe a variety of known or hypothesized classes of
transients.

\subsection{Radio Pulses from Ultra-high Energy Cosmic Rays or
 	Neutrinos}\label{sec:uhe} 

Intense, short-duration pulses ($\sim 1$~MJy in $\sim 10$~ns) at
decimeter and meter wavelengths have been detected from the
impact of ultra-high energy cosmic rays on Earth's atmosphere
\citep[e.g.,][]{jf65,w01,fg03},
and in at least one case even imaged by a radio array not dissimilar
to the LWDA \citep{fab+05}.  High-energy neutrinos impacting the lunar
regolith should also produce radio pulses via the Askar'yan effect
\citep{dz89,fg03}, though no such pulses have been detected\footnote{
No pulses from the Moon have been detected, but the Askar'yan effect
has been demonstrated in terrestrial accelerators
\citep{gsf+05,gbb+07}.
} from the
\objectname{Moon} to date
\citep{heo96,ghl+04,bdzko05,lbhm09,sbb+09,bsb+10,jea+10,jmg10}.  In the
case of the lunar neutrino radio pulses, the expected pulse amplitude
is more poorly constrained because it depends, in part, upon
properties (primarily surface roughness) of the lunar surface
\citep{a-mmvz06,sbb+06,jp09}.  Nonetheless, if sufficiently energetic
neutrinos exist, it is plausible that radio pulses would be emitted at
frequencies relevant for an LWDA-like instrument.

The intrinsic width of these radio pulses is extremely short, far
shorter than the 51~ms data acquisition time for an individual
baseline of the \hbox{LWDA}; as an example, \cite{fab+05} measure a
width narrower than 30~ns.  As noted above (\S\ref{sec:all-sky}),
during the commissioning of the
\hbox{LWDA}, a sporadic M\&C issue was recognized in that occasionally
a baseline would produce an anomalously large amplitude.  Upon
inspection, we found that baselines with anomalously large amplitudes
also had a phase of~0\arcdeg\ (which is one of the reasons that we
explain these anomalous amplitudes as being due to an M\&C issue).
Further, unlike the LOFAR Prototype Station
\citep[\hbox{LOPES},][]{fab+05}, we had no co-located particle detector to use as a
trigger for cosmic ray detection.  Subsequently, we employed a simple
threshold test, in which baselines with anomalous amplitudes and
phases near zero were
excised.  A consequence of this flagging methodology is that radio
pulses from high-energy particles at the zenith would not be recognized by our
analysis of LWDA data.

\subsection{Stars and Substellar Objects}\label{sec:star}

The \objectname[]{Sun} generates intense emission at LWDA frequencies,
notably Type~III and~IV radio bursts, but even the strongest such
radio bursts are far too faint to be detected over interstellar
distances with the current most sensitive meter- or
decameter-wavelength telescopes.  \cite{g86} considered the detection
of solar-type stars at low radio frequencies and finds that
millijansky sensitivities would likely be required to detect the
equivalent of the most intense solar radio bursts from nearby stars.
There are approximately 20 solar-type stars in the solar neighborhood
\citep[$< 10$~pc,][]{hna09}.  The absence of any transients that could
be identified with these stars puts only modest constraints on their
levels of activity.  Not only is the LWDA sensitivity likely to be
insufficient, if these stars have solar-like cycles over decadal time
scales, the absence of any flaring could merely reflect that a star is
currently in a quiescent phase of the cycle (much like the
\objectname[]{Sun} is in the current solar cycle).

Late-type stars (M dwarfs) and brown dwarfs also exhibit flaring
activity, particularly at centimeter- and meter wavelengths
\citep[$\gtrsim 300$~MHz, e.g.,][]{g02,hbl+07}.  Little is known about
the emission from brown dwarfs at LWDA frequencies.  However, the most
sensitive searches for decameter-wavelength emission from late-type
stars have, at best, marginal detections, even when those stars were
observed to be flaring at shorter wavelengths \cite{jkk90}.  Even
these marginally detected flares have flux densities of order 1~Jy.
While more numerous and likely to be closer than solar-type stars, the
LWDA sensitivity also means that we can place only modest limits on
the likelihood of strong flares from nearby M dwarfs.

Finally, within the solar system, in addition to solar radio bursts,
\objectname{Jupiter} also produces intense radio bursts due to an
electron cyclotron maser within its magnetosphere \citep{m05}.  This
emission cuts off above around~40~MHz, a frequency determined by the ratio
between the cyclotron and plasma frequencies as a function of altitude
within the Jovian magnetosphere.  

The magnetospheric emissions from solar system planets and the
discovery of extrasolar planets have motivated a number of both
theoretical
\citep{zqr+97,fdz99,ztrr01,flzbdr04,lfdghjh04,s04,gmmr05,z06,z07,gzs07,gpkmmr07}
and observational work
\citep{yse77,wdb86,bdl00,lfdghjh04,rzr04,gs07,lf08,scgjlb09} on
magnetospheric emissions from extrasolar planets, including some
before the confirmed discovery of any extrasolar planets.
Unfortunately, even the most optimistic predictions for extrasolar
planetary radio emission do not predict flux densities in excess
of~1~Jy at frequencies near~100~MHz.  Moreover, the most sensitive
searches for extrasolar planetary radio emission near~74~MHz place
sub-Jansky constraints on this emission \citep{lf08,l+09}.

\subsection{Pulsar Giant Pulses}\label{sec:giant}

We consider giant pulses from radio pulsars as an exemplar of a
short-duration, intense radio transient that, in principle, could be
detected throughout the Local Group by the LWDA \citep{cbhmk04}.  For the specific
case of the \objectname[]{Crab pulsar}, giant pulses have been seen at
frequencies as low as 74~MHz \citep{cclst69,rccrc70}, and \cite{sbhml99}
showed that the pulses are indeed broadband, at least over the
frequency range 610--4900~MHz.

The spectral index for \objectname[]{Crab pulsar} giant pulses is
steep ($\alpha \sim -3.4$, $S_\nu \propto \nu^\alpha$), and current
observations find no upper limit to the amplitude of giant pulses
\citep{sbhml99,lcumlf95}.  However, by comparison to lower frequency
observations, \cite{lcumlf95} also find that the rate of giant pulses
decreases with decreasing observational frequency, such that giant
pulses at~800~MHz are 400 times more frequent than at~146~MHz.
Scaling both the rate of giant pulses and the pulse amplitudes to the
LWDA frequency of~73.8~MHz, we estimate that, on average, the
\objectname[]{Crab pulsar} will produce a giant pulse once every
50~hr with a flux density exceeding 1~MJy (i.e., $\sim 2000\sigma$ for
the \hbox{LWDA}, Table~\ref{tab:simulate}).

At higher frequencies, \objectname[]{Crab pulsar} giant pulses have
extremely narrow widths, often being unresolved in time.  With a pulse
period of approximately 33~ms, even if the \objectname[]{Crab pulsar}
or a pulsar like it produced a sufficiently strong pulse to be
detected, only a single baseline would appear to be illuminated by our
data acquisition procedure (51~ms per baseline, \S\ref{sec:all-sky}).
However, at LWDA frequencies, pulse broadening becomes significant.
Pulse broadening increases the duration of the pulse, at the cost of
decreasing its amplitude.

In addition to being frequency dependent, pulse broadening also
depends upon direction through the Galaxy.  Using the NE2001 model
\citep{cl02}, we estimate that the magnitude of pulse broadening can
exceed 500~ms ($\sim 10$ LWDA baselines being illuminated) for
distances of order 5~kpc through the Galactic disk, implying Galactic
latitudes $|b| \lesssim 5\arcdeg$.  Unfortunately, these low Galactic
latitudes constitute only a small fraction of the total sky ($\lesssim
1\%$), and we also blanked the Galactic plane as part of our efforts
to remove strong sources from the images before conducting the
statistical tests.  Thus, we conclude that while giant-pulse emitting pulsars
seem capable of producing sufficiently strong pulses to be detectable,
either our serial data acquisition procedure would have not been
sensitive to such pulses or significant pulse broadening would have
rendered them too faint to be detectable. 

\subsection{Gamma-ray Bursts (GRBs)}\label{sec:grb}

We conclude this section by considering radio pulses from GRBs, as an
exemplar of possible radio pulses of extragalactic origin.
\cite{uk00} and \cite{sw02} both have predicted that GRBs
should have associated prompt emission, most likely below~100~MHz.
Predicted flux densities are highly model dependent and range from
essentially undetectable to in excess of~1~MJy.

Observationally, there have been a number of searches for radio pulses
associated with GRBs.  \cite{cmmcis81}, \cite{ismm82},
and \cite{alv89} all detected some dispersed radio pulses, but found
no convincing associations with GRBs.
\cite{b99} found a dispersed radio pulse apparently coincident
with \objectname{GRB~980329}, but it was narrowband, which has led to it being
interpreted as due to terrestrial interference.  Various searches for
radio pulses associated with GRBs (including precursor
pulses) have been conducted at~151~MHz
\cite{kgwwp94,kgwwp95,dgw+96}.
Typical upper limits have been approximately 100~Jy.

\subsection{Exotica}\label{sec:exotica}

One of the motivations for this, and similar, searches is simply that
previously unknown classes of sources may be discovered.  While it is
difficult to assess the probability of an unknown class of sources, we
can consider extraterrestrial transmitters \citep{cm59} as an example
of an exotic population of sources.\footnote{
Doing so is also appropriate given that this year is the
$50^{\mathrm{th}}$ anniversary of the first search for radio
transmissions, F.~Drake's Project \hbox{OZMA}.
}

As \cite{lz07} discuss, much of the human-generated radio radiation
from the \objectname{Earth} is emitted in the range 50--400~MHz.
Indeed, the upper end of the LWDA's operational frequency range was
chosen to avoid FM radio broadcasts, and \S\ref{sec:detect}
illustrates the detection of TV signals with the \hbox{LWDA}.  In
contrast to \cite{lz07}, however, we consider short duration pulses,
such as might originate from a transmitter on a planet rotating into
or out of view, whereas they consider detecting signals in long
integrations ($\sim 1$~month).

The LWDA's frequency sub-bands of~20~kHz are well matched to the
typical bandwidths, at least for terrestrial transmitters in this
frequency range.  There is an increasing usage of so-called spread
spectrum transmitters, with bandwidths (much) larger than 20~kHz.
However, these are often at higher frequencies, and an LWDA-like
instrument would still receive all of the power within its received
bandwidth.

Unfortunately, with a 500~Jy sensitivity, it is a relatively simple
matter to show that a transmitter at a distance of~10~pc would have to
have an effective isotropic radiated power (EIRP) of about $10^{17}$~W
(170~dBW) to have been detectable with the \hbox{LWDA}.  By contrast,
the EIRP of even powerful military radars, such as the Air Force Space
Surveillance System (AFSSS), only approach $5 \times 10^{13}$~W
(137~dBW).  Further, given that all (or most) of the stars
within~10~pc are cataloged, it is not clear that an all-sky search,
such as the one pursued here, is the most effective search strategy,
unless one postulates a set of freely floating ``beacons.''

\subsection{General Considerations}\label{sec:general}

Examination of Figure~\ref{fig:phasespace} shows that an LWDA-like
instrument might be able to probe a significant region in this phase
space, and previous sections have discussed a range of source classes
that the LWDA might have been able to detect, in principle, even
though the implementation of the data acquisition system resulted in no
detections.  In addition to the aforementioned issue that a more rapid
time sampling would likely be required, are there other considerations
for a dipole phased array?

Much of the phase space accessible, or potentially accessible, to an
LWDA-like instrument is for \emph{coherent} transients, nominally
taken to be those with brightness temperatures in excess
of~$10^{12}$~\hbox{K}.  Such transients will necessarily evolve
rapidly.  \cite{cm03} have discussed searches for fast radio
transients.  A key consideration for such searches is the effective
time resolution, which is determined not only by the instrumental
characteristics (receiver bandpass) but also by dispersion smearing
and radio-wave scattering.

Following \cite{cm03}, we consider two time scales related to the time
resolution, and implied processing, for searches with an LWDA-like
instrument.  The highest time resolution that the instrument would be
able to obtain is simply
\begin{equation}
\Delta t_{\Delta\nu} \sim (\Delta\nu)^{-1},
\label{eqn:dt_dn}
\end{equation}
for a receiver frequency channelization or sub-band of~$\Delta\nu$, 
which for the LWDA ($\Delta\nu = 20$~kHz) was 50~$\mu$s.  Because of
the dispersion relation of the interstellar medium, probing to any
distance for fast pulses requires de-dispersion.  As an illustration,
we consider how far one could probe, ignoring dispersion.  We take as
a criterion that the dispersion across the receiver sub-band produces
a smearing comparable to time resolution implied by
Equation~(\ref{eqn:dt_dn}).  Then, from \cite{cm03}, 
\begin{equation}
\delta\mathrm{DM}
 = 120\,\mathrm{pc}\,\mathrm{cm}^{-3}\,(\Delta\nu_{\mathrm{kHz}})^{-2}\nu_{\mathrm{MHz}}^3,
\label{eqn:dt_ddm}
\end{equation}
where $\Delta\nu_{\mathrm{kHz}}$ is the receiver sub-band in units of
kHz, and $\nu_{\mathrm{MHz}}$ is the central frequency in units of
MHz, and the error in the dispersion measure~$\delta\mathrm{DM}$ is in
the canonical units of~pc~cm${}^{-3}$.  Alternately,
$\delta\mathrm{DM}$ can be considered the increment in dispersion
measure (DM) for searches
that incorporate searching through dispersion.  For the \hbox{LWDA},
$\delta\mathrm{DM} = 0.1$~pc~cm${}^{-3}$, implying that a search for
fast transients with the LWDA that did not incorporate de-dispersion
could only probe to an effective DM of this value.  For reference,
\objectname{PSR~J0108$-$1431} has the lowest known \hbox{DM},
2.38~pc~cm${}^{-3}$, at a distance of~240~pc \citep{mhth05,dtbr09}.

Conducting a search that incorporated compensation for dispersion
smearing would effectively mean conducting a de-dispersion search on
each resolution element (beam) within the field of view \citep{jimcordes09}.
In contrast to proposed searches at higher frequencies ($\sim 1$~GHz),
the much lower resolution at LWDA frequencies makes this a less
onerous task.  For instance, with the LWDA's resolution of~12\arcdeg,
there are only about 150 independent resolution elements in the entire
sky.  Further, searching each is of course an embarrassingly parallel
problem.

\section{Conclusions}\label{sec:conclude}

We have described the Long Wavelength Demonstrator Array (LWDA) and
its operation as an all-sky transient monitor.  A 16-element dipole
phased array, operating over the frequency range 60--80~MHz, the
LWDA was used as a technical test bed for the \hbox{LWA}.  We have described the signal flow from the antennas to the
receivers and to a simple software correlator as well as the
configuration of the dipoles.  

The individual dipoles of the LWDA had a field of view comparable to
the entire sky.  In late 2006 and early 2007, we used the LWDA as a
transient search instrument by making a series of all-sky images at 
frequencies of~61 and~73.8~MHz.  The 61~MHz observations were designed
explicitly to search for reflections from ionized meteor trail during
the Leonid meteor shower in~2006.  The 73.8~MHz observations were
designed to search for astronomical transients, and we acquired a
total of~106~hr of data, with a time sampling ranging from~2
to~5~minutes between images.  We were able to detect solar flares and,
by utilizing a special-purpose mode, TV reflections off ionized meteor
trails during the 2006 Leonid meteor shower.  We detected no radio
transients outside of the solar system above a flux density limit
of~500~Jy, equivalent to a limit of no more than about $10^{-2}$
events~yr${}^{-1}$~deg${}^{-2}$, having a pulse energy density $\ga
1.5 \times 10^{-20}$~J~m${}^{-^2}$~Hz${}^{-1}$ at~73.8~MHz for pulse
widths of about~300~s.  This event rate is comparable to many existing
limits from previous all-sky surveys, but at a lower frequency than
most previous all-sky searches.

There are a number of emerging arrays (e.g., \hbox{MWA}, \hbox{LOFAR},
\hbox{LWA})
for which all-sky, or at least very wide-field, imaging will be a
capability.  
All of these arrays will offer at least two advantages with respect to
the \hbox{LWDA}.  First, all are anticipated to be operational for a
far longer time than was the \hbox{LWDA}, which should improve upon
the event rates substantially.  Second, all of these arrays will have
a higher angular resolution imaging capability, so that if a transient
were to be detected, much higher precision astrometric information
could be obtained.

\acknowledgements

The LWDA was a joint project of \hbox{NRL}, the Applied Research
Laboratories of the University of Texas at Austin, and the University of
New Mexico.
We thank the members of the University of New Mexico faculty and students who
assisted with the infrastructure, construction, and maintenance of the LWDA and the many staff members of the NRAO who provided technical
support in establishing the LWDA near the \hbox{VLA}.  We thank
K.~Weiler, D.~Munton, and L.~J~Rickard for their guidance during
various stages of the project.  We thank J.~Cordes for providing the
software to generate Figure~\ref{fig:phasespace}.
This research has made use of NASA's Astrophysics Data System.  
The \anchor{http://lunar.colorado.edu}{LUNAR consortium}, headquartered at 
  the University of Colorado, is funded by the NASA Lunar Science 
  Institute (via Cooperative Agreement NNA09DB30A) to investigate 
  concepts for astrophysical observatories on the Moon. 
Basic research in radio astronomy at the NRL is supported
by 6.1 Base funding.

\end{document}